\documentclass[aps,prl,twocolumn,groupedaddress,showpacs]{revtex4-1}

\usepackage{graphicx,color}
\usepackage{epsfig}
\usepackage{dcolumn}
\usepackage{amsmath,amssymb,bm}
\usepackage{hyperref,url}
\usepackage[noabbrev]{cleveref}
\usepackage{CJK}
\usepackage{lineno}
\usepackage[english]{babel}
\usepackage[T1]{fontenc}
\usepackage{pifont}
\usepackage{amsfonts}
\usepackage{wasysym}
\usepackage{amsbsy}
\usepackage{psfrag}
\usepackage{color}
\usepackage{multirow}
\usepackage{cases}
\usepackage{stackengine}
\usepackage{float}
\usepackage{blkarray}
\usepackage{cancel}
\usepackage{mathrsfs}
\usepackage{empheq}
\usepackage{mathptmx} 

\begin{document}

\preprint{APS/123-QED}

\title{\textbf{Self-scalings of bubble pinch-off and inner jet emitting in a tapered co-flow} 
}%

\author{B. J. Ruan}
\author{{Z. L. Wang}}%
 \email{wng_zh@i.shu.edu.cn}
\affiliation{Shanghai Institute of Applied Mathematics and Mechanics, Shanghai Key Laboratory of Mechanics in Energy Engineering, Shanghai University, Yanchang Road 149, Shanghai, 200072, P.R. China
}%

\date{\today}

\begin{abstract}

This study systematically investigates the formation mechanisms and scaling laws governing sub-millimeter bubble-jet generation in convergent coaxial microchannels. Experimental observations reveal four distinct evolutionary stages in monodisperse bubble formation: growth, necking, detachment, and stabilization. Through necking rupture dynamics modeling, we demonstrate that the coupled effects of nozzle insertion length ($x$) and convergence angle ($\alpha$) dominate neck width evolution, exhibiting universal power-law scaling across geometric configurations. Crucially, the spatiotemporal evolution of necking interfaces under shear demonstrates self-similarity: temporal evolution follows a power-law decay with respect to remaining time ($T-t$), while spatial scaling correlates with characteristic dimension $W_\text{local}$ through power-law relationships. Multi-stage mathematical models successfully describe this self-similar interfacial behavior, confirming their predictive validity. Furthermore, we identify the core mechanism of bubble-jet formation as dynamic modulation through convergent zone flow fields characterized by front-end stretching and rear-end squeezing interactions. Experimental validation shows consistent jet velocity evolution across flow regimes under multiphysics coupling, establishing geometric similarity principles governing flow structures and dynamic behaviors in convergent microchannel architectures.
 
\end{abstract}


\pacs{47.55.Dz 47.62.+q 47.55.Kf}              
\maketitle

\section{Introduction}
\label{Introduction}

Sub-millimeter bubble jets are a unique and intriguing phenomenon with significant potential in various applications, including drug delivery, biochemical reactions, and cell puncture\cite{rodriguez2015generation,patel2021advances,nakagawa2017pulsed}. These jets are typically induced through external excitations such as ultrasound, shock waves, or lasers, which trigger cavitation and subsequent jet formation\cite{yusof2016physical,nikolov2019air,tan2009interfacial,deng2024impact,huang2021physical,luo2019jet}. However, the complexity of the applied conditions often results in limited stability and controllability, hindering their practical implementation.

In contrast, research on millimeter-scale bubble jets is more prevalent, focusing on phenomena such as bubble emergence, coalescence, and stretching\cite{blanco2020sea,lorenceau2004air,liger2003capillary,cochran2017sea,chingin2018enrichment}. These studies are often related to natural processes like oceanic atomization, aerosol generation, atmospheric bubble dynamics, cloud formation, and air pollution control\cite{cheng2020numerical,berny2020role,seon2012large}. In such macroscopic scenarios, gravity plays a dominant role, and the phenomena typically occur in open environments with free boundaries.

At the sub-millimeter scale, however, the dynamics are governed by capillary effects rather than gravity. The characteristic length scale is defined by the capillary length ($l_c=\sqrt{\sigma/(\rho g)}\approx 2.7,\text{mm}$), and the Bond number ($Bo=\rho g L^2/\sigma$) is usually less than unity\cite{brasz2018minimum,cheng2020numerical,berny2020role}. Under these conditions, surface tension dominates, and capillary forces significantly influence gas-liquid deformation. Consequently, sub-millimeter bubble jets often require external stimuli to overcome these constraints. For instance, ultrasound induces bubble oscillation and cavitation through periodic pressure waves\cite{nikolov2019air,tan2009interfacial}, while shock waves drive rapid compression and expansion through abrupt pressure changes\cite{luo2019jet}. Similarly, lasers generate rapid internal pressure differences via the photothermal effect\cite{brown2011time,patrascioiu2014laser}.

In the realm of digital microfluidics, where Reynolds numbers are typically less than unity ($Re < 1$), bubble-jet phenomena are rarely observed\cite{ganan2017revision}. A notable exception is the inward-rolling bubble-jet at relatively high Reynolds numbers ($Re=\frac{\rho_c u_c d}{\mu}\approx 400$)\cite{dasouqi2022effect}. However, the generation of sub-millimeter bubble jets in digital microfluidic systems remains unexplored, presenting a promising avenue for future research.

The interfacial evolution dynamics during the slug flow necking stage in microchannels exhibits intrinsic correlations with bubble-jet excitation processes\cite{deike2018dynamics,andredaki2021accelerating}. The exponential decay characteristics of necking width directly govern the initial perturbation wavelength of bubble-jet formation. Current understanding remains limited regarding the regulatory mechanisms of shear force gradients versus surface tension competition on jet excitation thresholds during bubble necking in convergent microchannels. This study systematically investigates necking width attenuation characteristics through experimental observation, elucidates intrinsic relationships governing necking dynamics, and establishes a predictive model for necking gas-liquid interface profiles.

Through an in-depth study of the generation mechanisms of sub-millimeter bubble jets, this research offers new insights for the design and application of digital microfluidic systems, expanding their potential applications in biomedical and chemical engineering fields.

\section{Experimental Set-up}

\begin{figure}[!t]
  \centering
  \includegraphics[width=8.5cm,keepaspectratio]{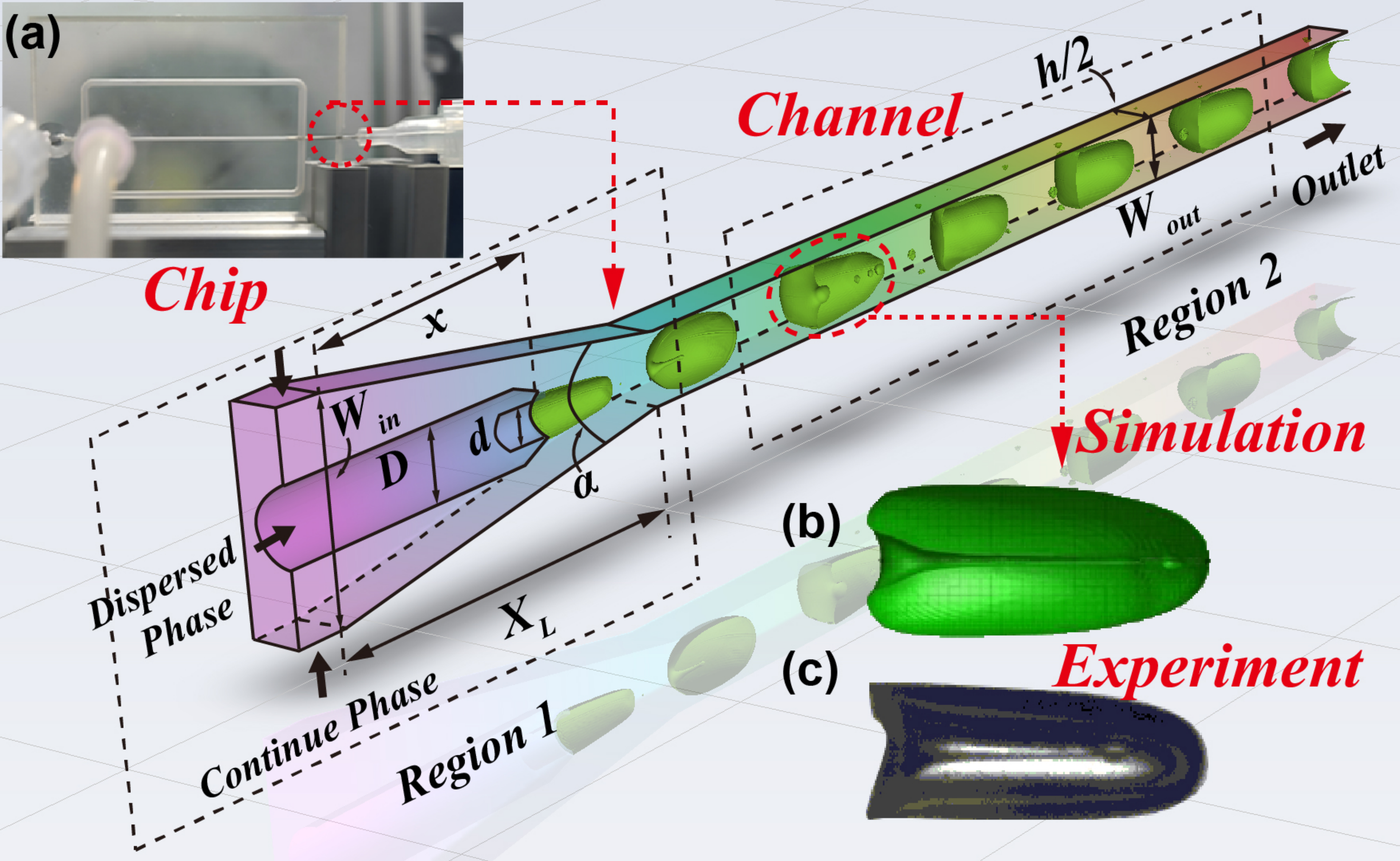}
  \caption{  
  Bubble-jets generating in tapered chips.(a) microchip on-line (top left) and sketch of internal jet formation within the gas-liquid two-phase flow in the tapered channel (middle). Region 1 is the main tapered zone,  and the inlet exhibits a rectangular cross-section ($h\times W_{\text{in}}$, where $h = 400\mu\text{m}$ and $W_{\text{in}} = W_{\text{out}} + 2X_L\tan(\alpha/2)$). The length of the tapered region is $X_L = 2600\mu\text{m}$. The outer diameter of the needle is $D = 400\mu\text{m}$, and the inner diameter is $d = 200\mu\text{m}$. The convergence angles are $\alpha = 5^{\circ}, 11^{\circ}, 17^{\circ}, 29^{\circ}$, and the needle displacements are $x = 500\mu\text{m}, 1000\mu\text{m}$, and $1500\mu\text{m}$. Region 2 is the parallel pipe flow  with a cross-section of $h\times W_{\text{out}} = 400\times 400\mu\text{m}^{2}$.(b) A typical image of sub-millimeter bubble jetting from 3D Volume of Fluid (VoF) simulations.(c) Experimental image corresponding to (b).
  }
   
  \label{fig:Schematic}
\end{figure}

This letter reports our inventions of generating digital sub-millimeter bubble-jet, as well as relevant findings and laws. We designed polymethylmethacrylate (PMMA) micro-chips with tapered channels \cite{wang2015speed,wang2022universal}, as shown in Fig.\ref{fig:Schematic}(a).The manufacturing process involves three sequential stages: 1) Substrate preparation through material selection and pretreatment including precision cutting, multi-stage purification (deionized water rinsing, alkaline solution cleaning, and acetone ultrasonic treatment); 2) Micro-channel fabrication via a precision engraving machining center; 3) Thermal bonding assembly comprising fixture preheating, adhesive activation at critical temperature under constant pressure, followed by cooling stabilization. The functionalized chips were subsequently subjected to performance evaluation focusing on interfacial dynamics. Significantly, the fabricated microchips enabled precise regulation of unstable gas-liquid co-axial flow, yielding distinct flow patterns including slug, jetting, dripping, and annular regimes through controlled parameter modulation. The existence of the tapered zone breaks down the Galilean invariance to give two geometry variables: the taper angle $\alpha$ and the relative inner needle displacement $x$. In this work, $\alpha = 5^\circ, 11^\circ, 17^\circ, 29^\circ$, and $x = 500,1000$, and $1500 \mu\text{m}$, respectively. Their combinations yield 12 experimental data sets.The gas (argon) is injected as the dispersed phase from the middle needle, while the continuous phase (deionized water) is injected from the outer channel. The flow rates of the continuous phase (deionized water) $Q_c$ and the dispersed phase (argon) $Q_d$ provided by a peristaltic pump and a mass flow controller range of 5 $\sim$ 100 mL/min and 5 $\sim$ 50 mL/min. The interfacial tension $\sigma$ is 72.8 mN/m. The density $\rho_c$ and viscosity $\mu_c$ of water are 986.2 kg/m$^3$ and 1.23 mPa$\cdot$s, respectively, and the density $\rho_ d$ and viscosity $\mu_d$ of argon are 1.613 kg/m$^3$ and 0.022 mPa$\cdot$s (measured at room temperature).

The dynamics of the bubbles and jets are illuminated by a high-intensity LED light, and snapshots are captured at 20,000 Hz (Phantom V611-16G-M), with an exposure time of 10 $\mu s$ and a resolution of $1280\times240$ pixels. The dynamic process of jet excitation within monodisperse bubbles can be clearly captured, as shown in Fig.\ref{fig:Schematic}(c). We also used the Volume of Fluid (VoF) to compare and verify the corresponding experimental results (Fig.\ref{fig:Schematic}(b)).

\section{Bubble Formation Dynamics}

\begin{figure*}[!t]
  \centering
  \includegraphics[width=15cm,keepaspectratio]{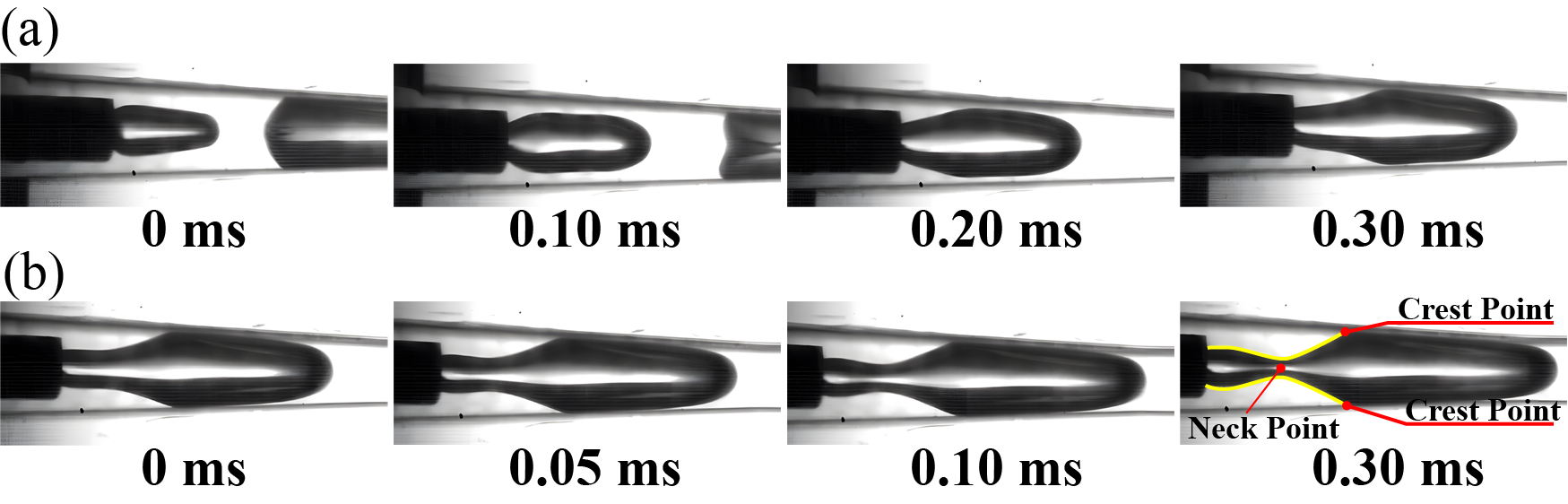}
  \caption{Experimental snapshots of bubble formation process: (a) growth stage; (b) necking stage.}
  \label{fig:1-气泡流型生长阶段和颈缩阶段}
\end{figure*}

Based on the aforementioned classification of bubble-jet flow patterns within the main channel and the theoretical framework of dual critical conditions, this chapter further conducts systematic visualization experiments employing high-speed microscopic imaging technology to investigate the generation dynamics of dispersed-phase bubbles in convergent microchannels. As illustrated in \ref{fig:1-气泡流型生长阶段和颈缩阶段}, the bubble evolution process at the nozzle outlet exhibits four distinct kinetic phases with characteristic interfacial dynamics: (I) Growth Stage, (II) Necking Stage, (III) Detachment Stage, and (IV) Stabilization Stage. These sequential stages are governed by unique interfacial evolution mechanisms coupled with hydrodynamic interactions, collectively dictating the morphological transformation and transport characteristics of bubbles throughout their lifecycle.

\begin{figure*}[!t]
  \centering
  \includegraphics[width=15cm,keepaspectratio]{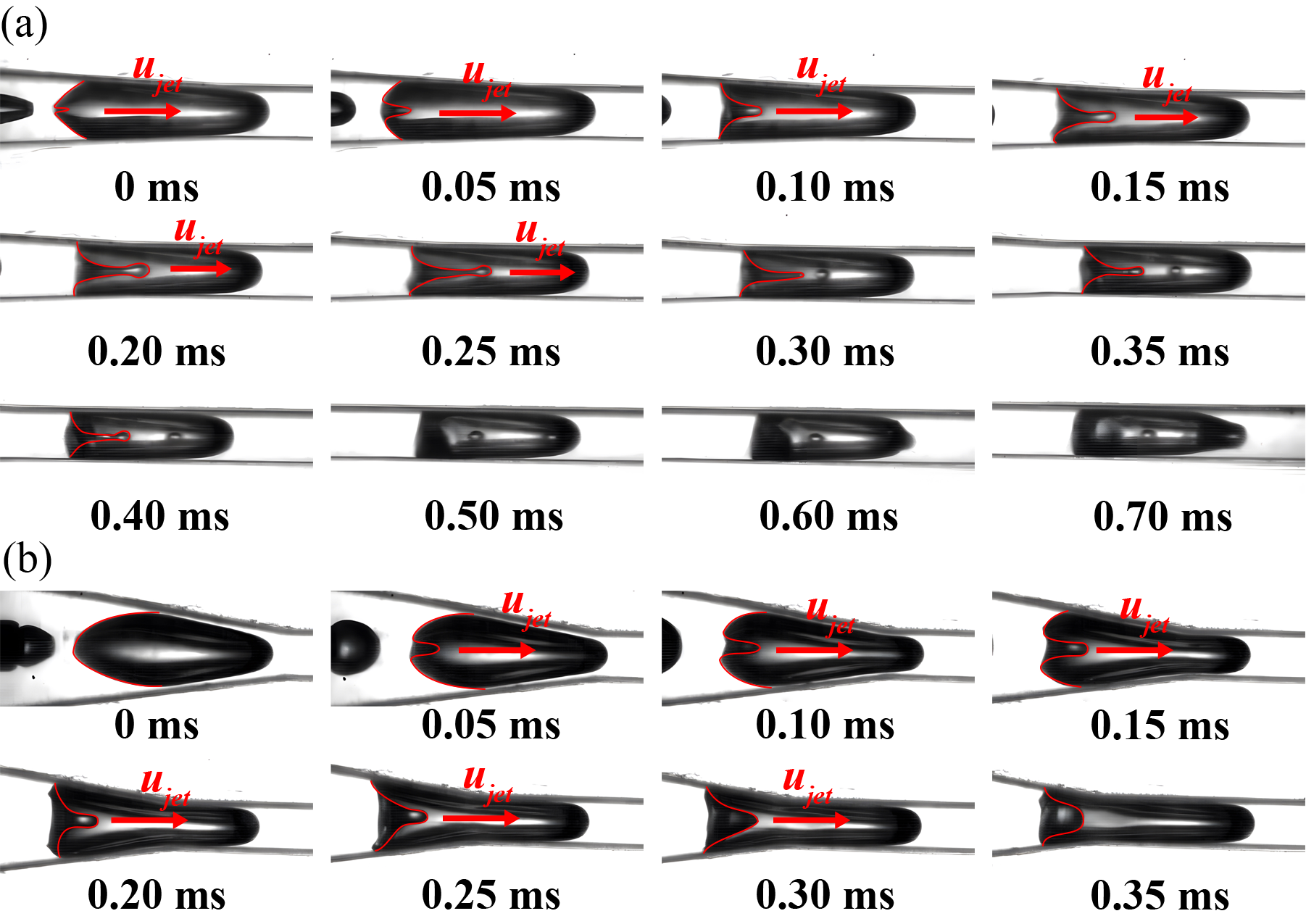}
  \caption{Experimental snapshots of bubble detachment stage: (a) Inner Jet without Droplet; (b) Inner Jet with Droplet.}
  \label{fig:2-气泡流型脱离阶段}
\end{figure*}

The growth stage, as depicted in Fig. \ref{fig:1-气泡流型生长阶段和颈缩阶段}(a), exhibits characteristic dynamics during the time interval of 0-30 ms. Within this temporal regime, the bubble initiates its protrusion from the nozzle exit with quasi-ellipsoidal morphology, demonstrating monotonic volumetric expansion governed by the dispersed-phase supply rate, continuous-phase volumetric flow rate, interfacial tension, and fluid viscosity. Owing to the diminished inertial effects inherent to microfluidic confinement, the equilibrium between viscous dissipation and gas expansion pressure governs the dynamics, resulting in sustained growth until reaching critical conditions defined by the Rayleigh-Plateau instability threshold.

The necking stage emerges as the bubble undergoes progressive necking at the nozzle interface due to hydrodynamic focusing effects, as captured in Fig. \ref{fig:1-气泡流型生长阶段和颈缩阶段}(b). Under the geometric confinement of microfluidic channels, the bubble develops a characteristic prolate-pear morphology with axisymmetric necking features. This stage is fundamentally governed by the competition between interfacial tension and viscous shear stresses: The former acts to preserve interfacial continuity through energy minimization, while the latter generates an extensional flow field that induces progressive thinning of the neck region. Such hydrodynamic instability ultimately triggers flow pattern transition through capillary-driven singularity formation, consistent with modified Rayleigh-Plateau instability mechanisms in confined geometries.

During the bubble detachment phase, as illustrated in Fig. \ref{fig:2-气泡流型脱离阶段}(a), the necking process reaches its critical point where the bubble-nozzle connection undergoes abrupt rupture at $t = 0$ ms, forming a discrete entity. The instantaneous disappearance of surface tension and gas momentum flux disrupts force equilibrium, resulting in an inverted conical morphology at the bubble posterior. At $t = 0.05$ ms, constrained growth at the bubble front due to microchannel wall confinement induces localized collapse. This collapse propagates along the posterior interface, generating capillary wave trains that drive substantial deformation followed by gradual relaxation.  

Concurrent inward folding of the posterior interface initiates liquid jet formation. Between $t = 0.10$ ms and $t = 0.25$ ms, convergent flow at the bubble posterior undergoes inertial acceleration, developing a slender jet structure. Progressive thinning of the jet diameter occurs, where morphological characteristics and velocity profiles are governed by bubble dimensions, liquid velocity fields, and fluid properties.  

The subsequent evolution ($t = 0.30$ ms to $t = 0.40$ ms) demonstrates jet tip narrowing with potential droplet pinch-off events under compressive stresses. These satellite droplets exhibit intra-bubble migration, forming encapsulated droplet-bubble complexes. Final stages ($t = 0.50$ ms to $t = 0.60$ ms) reveal high-velocity jet/droplet penetration towards the bubble anterior, generating surface protrusions or cusp formations. Measured kinetic energies of these transient structures significantly exceed the baseline translational velocity of pressure-driven monodisperse bubbles.

As shown in Fig. \ref{fig:2-气泡流型脱离阶段}(b), although the jet velocity remains insufficient to achieve droplet pinch-off or neck rupture, the maximum elongation length of this intra-bubble jet is clearly observable. Termed as an inertially stretched jet induced by bubble surface deformation, it exhibits an inverted conical morphology with a tapered trailing edge, analogous to gravity-driven large-scale bubble-jet rupture in unconfined environments. However, the characteristic dynamics of such submillimeter bubble-jets under microchannel confinement has not been systematically investigated.

Upon entering the non-convergent region of the microchannel during the stabilization phase, the bubble achieves morphological and kinematic stabilization. Its geometry remains quasi-stationary while the translational velocity equilibrates with the surrounding flow field. Within the laminar flow regime characteristic of microfluidic systems, this stage exhibits steady-state motion dynamics with minimal interfacial fluctuations.

\section{Necking Stage Analysis}

\begin{figure*}[!t]
  \centering
  \includegraphics[width=15cm,keepaspectratio]{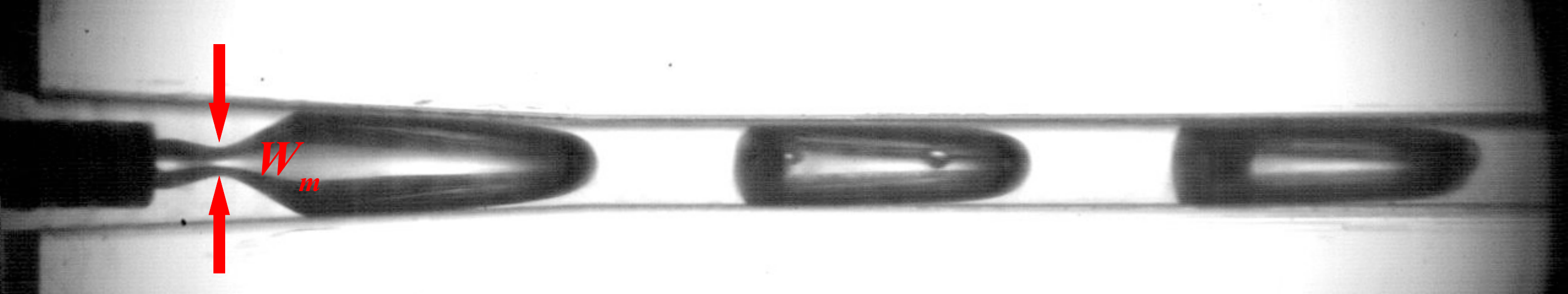}
  \caption{Schematic illustration of neck width evolution during bubble formation process.}
  \label{fig:3-气泡形成过程中颈部宽度示意图}
\end{figure*}

To quantitatively characterize bubble formation dynamics, we investigate the temporal evolution of neck width during bubble growth. During the initial phase of bubble growth, where the bubble exhibits an ellipsoidal morphology without discernible necking, a reference position must be defined to establish the necking region. The neck width $W_\text{m}$ is operationally defined as the minimum transverse dimension at this designated necking position, as schematically illustrated in Fig. \ref{fig:3-气泡形成过程中颈部宽度示意图}.

\begin{figure}[!h]
  \centering
 \includegraphics[width=8.5cm,keepaspectratio]{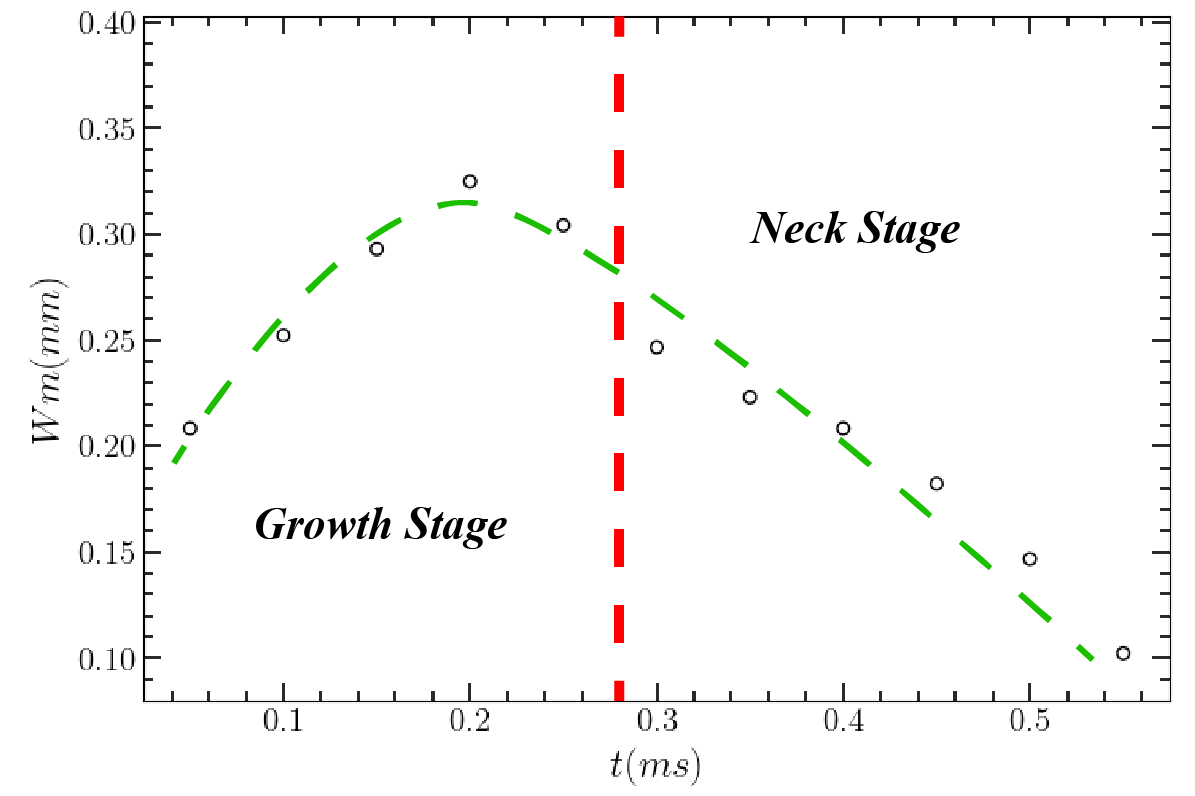}
  \caption{Temporal evolution of neck width during the bubble formation cycle.}
  \label{fig:4-气泡形成周期内颈部宽度随时间演变过程}
\end{figure}

Figure \ref{fig:4-气泡形成周期内颈部宽度随时间演变过程} presents the temporal evolution of neck width ($W_\text{m}$) during a complete bubble formation cycle in the gas-liquid two-phase flow system of Specimen A2. The data reveal two distinct regimes in monodisperse bubble formation: (1) Neck width expansion during the growth phase, and (2) progressive neck contraction preceding detachment. The phase transition from growth to detachment is marked by an abrupt decrease in $W_\text{m}$, with the critical neck width at rupture initiation slightly exceeding the nozzle inner diameter $d$.

\begin{figure}[!h]
  \centering
 \includegraphics[width=8.5cm,keepaspectratio]{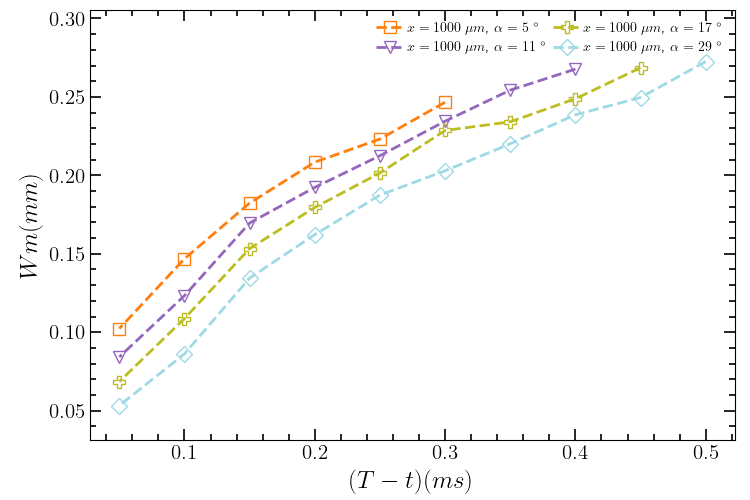}
  \caption{Relationship between bubble neck width and remaining time under identical convergence lengths \( x \) but varying convergence angles \( \alpha \).}
  \label{fig:5-不同式样气泡颈部宽度与剩余时间的关系(1)}
\end{figure}

\begin{figure}[!h]
  \centering
 \includegraphics[width=8.5cm,keepaspectratio]{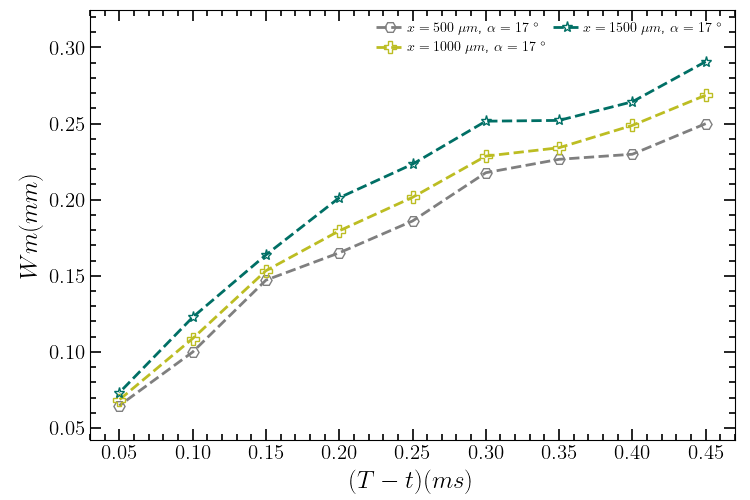}
  \caption{Relationship between bubble neck width and remaining time under fixed convergence angle \( \alpha \) but varying convergence lengths \( x \).}
  \label{fig:5-不同式样气泡颈部宽度与剩余时间的关系(2)}
\end{figure}

To characterize neck width evolution during the contraction phase, we define $T$ as the total formation cycle duration of slug flow and $t$ as the instantaneous time during bubble formation, where the remaining time is expressed as $(T-t)$. Figures \ref{fig:5-不同式样气泡颈部宽度与剩余时间的关系(1)} and \ref{fig:5-不同式样气泡颈部宽度与剩余时间的关系(2)} present the neck width ($W_\text{m}$) versus remaining time relationships under experimental conditions $Q_\text{c}=100$ ml/min (continuous phase flow rate) and $Q_\text{d}=20$ ml/min (dispersed phase flow rate). Figure \ref{fig:5-不同式样气泡颈部宽度与剩余时间的关系(1)} compares specimens with varying convergence angles ($\alpha$), while Figure \ref{fig:5-不同式样气泡颈部宽度与剩余时间的关系(2)} examines different convergence lengths ($x$).

\section{Necking Rupture Model}

To establish scaling laws governing bubble necking dynamics in convergent microchannel architectures, our analysis must simultaneously account for the coupled effects of convergence angle ($\alpha$) and axial convergence distance ($x$). This necessitates the introduction of a local characteristic dimension $W_\text{local}$ that encapsulates geometric confinement effects. The macroscopic scaling relationship can be expressed as:  
\begin{equation} 
W_\text{m}=F(\rho_\text{d}, \rho_\text{c}, \mu_\text{d}, \mu_\text{c}, \sigma; h, W_\text{out}, D, d, W_\text{local}; Q_\text{d}, Q_\text{c};(T-t))
\label{eq:W_m-F}
\end{equation}
The physical dependencies of neck width \( W_\text{m} \) can be systematically categorized into three parameter systems: physical properties parameters (\( \rho_\text{d} \), \( \rho_\text{c} \), \( \mu_\text{d} \), \( \mu_\text{c} \), \( \sigma \)), geometric parameters (\( h \), \( W_\text{out} \), \( D \), \( d \), \( W_\text{local} \)), and flow control parameters (\( Q_\text{d} \), \( Q_\text{c} \), \( T-t \)). Although dimensionality reduction of geometric features in the convergent region is achieved by introducing the similarity variable \( W_\text{local} \), and the correlation between \( W_\text{m} \) and \( W_\text{local} \) is established based on similarity principles, the high dimensionality of the parameter space still poses challenges in cost and complexity for full-parameter experimental validation. To address this, a power-law predictive model for \( W_\text{m} \) during the necking phase is proposed, following methodologies in prior studies \cite{fu2010scaling}.
\begin{equation}
W_\text{m} =k(T-t)^{\alpha} \label{eq:W_m}
\end{equation}

This study investigates the variation of minimum neck width \( W_\text{m} \) with remaining time \((T-t)\) under identical flow rates across different geometric configurations. In bubble necking rupture phenomena, the necking characteristics are primarily governed by local geometric features, fluid properties, and temporal parameters. To precisely describe the necking dynamics, a non-dimensional analysis is implemented, requiring dimensionality reduction through physical mechanism analysis. First, simplification of dispersed phase parameters: Experimental observations confirm that continuous phase (liquid) dominates interfacial dynamics during necking, with negligible mechanical contributions from dispersed phase (gas) density \( \rho_\text{d} \) and viscosity \( \mu_\text{d} \) due to their orders-of-magnitude differences (\( \rho_\text{d}/\rho_\text{c} \sim 10^{-3} \), \( \mu_\text{d}/\mu_\text{c} \sim 10^{-2} \)). Second, geometric parameter focusing: Global geometric parameters (\( h \), \( W_\text{out} \), \( D \), \( d \)) are encapsulated through the local characteristic width \( W_\text{local} \), uniquely determined by convergence angle \( \alpha \) and axial position \( x \), effectively describing geometric confinement in the necking region. Finally, implicit treatment of flow parameters: Flow rates \( Q_\text{d} \) and \( Q_\text{c} \) define the system's temporal scale \( T \), whose influence is inherently embedded in the remaining time \((T-t)\) under constant flow conditions. Through this theoretical framework, the governing variables of \( W_\text{m} \) are identified and simplified into the following functional form:
\begin{equation}
W_\text{m}=F(\rho_\text{c},\mu_\text{c}, \sigma; W_\text{local};(T-t)) 
\label{eq:two-variable system8}
\end{equation}
Consequently, the model is reduced to an expression dependent solely on continuous phase properties (\( \rho_\text{c} \), \( \mu_\text{c} \), \( \sigma \)), local geometry (\( W_\text{local} \)), and temporal scaling (\( T-t \)). This simplified formulation preserves the essential physics while significantly reducing the dimensionality of the experimental parameter space.

Through dimensional analysis, the density $\rho_\text{c}$, surface tension $\sigma$, and viscosity $\mu_\text{c}$ were selected as fundamental scaling parameters to construct dimensionless groups via the Buckingham $\pi$ theorem, enabling systematic non-dimensionalization of the governing equations:
\begin{equation}
La_\text{w} \sim f(La_\text{local},\tau)
\label{eq:two-variable system9}
\end{equation}
Three dimensionless groups were constructed: the necking Laplace number \( La_\text{w} = W_\text{m}\sigma\rho_\text{c}/\mu_\text{c}^2 \) characterizing the balance between viscous and capillary forces at the necking scale, the dimensionless remaining time \( \tau = (T-t)\rho_\text{c}\sigma^2/\mu_\text{c}^3 \) representing the temporal progression of necking, and the local geometric Laplace number \( La_\text{local} = W_\text{local}\sigma\rho_\text{c}/\mu_\text{c}^2 \) quantifying the confinement intensity induced by channel convergence.

\begin{figure}[!h]
  \centering
 \includegraphics[width=8.5cm,keepaspectratio]{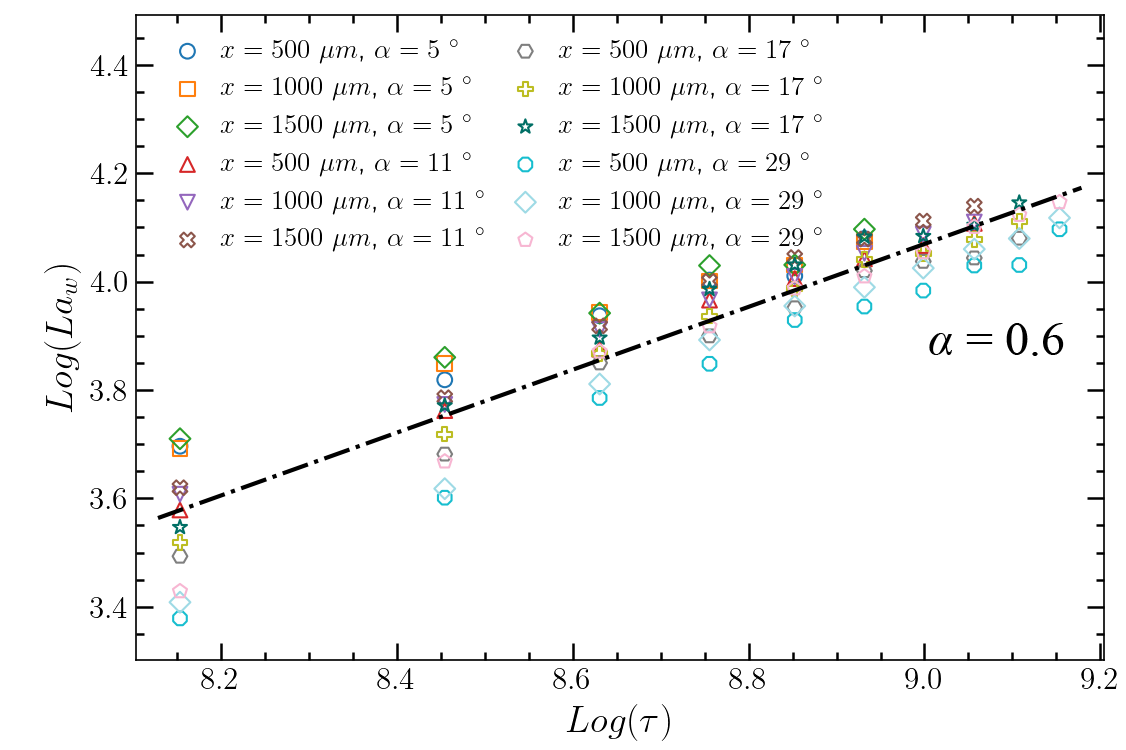}
  \caption{Relationship between the necking width number \( La_\text{w} \) and the remaining time number \( \tau \).}
  \label{fig:6-颈缩宽度和时间无量纲公式}
\end{figure}

\begin{figure}[!h]
  \centering
 \includegraphics[width=8.5cm,keepaspectratio]{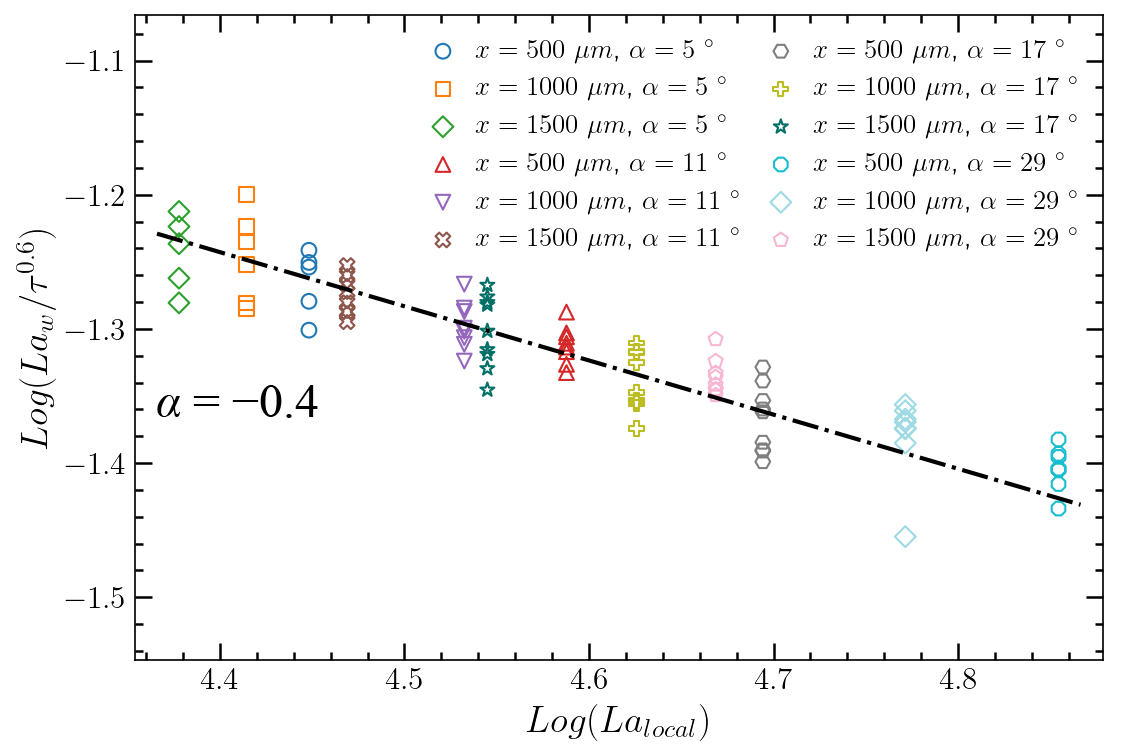}
  \caption{Relationship between the necking width number \( La_\text{w} \), remaining time number \( \tau \), and local geometric number \( La_\text{local} \).}
  \label{fig:7-颈缩宽度和局部宽度无量纲公式}
\end{figure}

Experimental analysis reveals functional relationships between these dimensionless groups, as demonstrated in Figures \ref{fig:6-颈缩宽度和时间无量纲公式} and \ref{fig:7-颈缩宽度和局部宽度无量纲公式}, where \( La_\text{w} \) exhibits a power-law dependence on \( \tau \) and \( La_\text{local} \), confirming the dominance of capillary-viscous interactions during necking dynamics.
\begin{equation}
La_\text{w}\sim  La_\text{local}^{-0.4}\cdot\tau^{0.6}
\label{eq:two-variable system10}
\end{equation}
\begin{figure}[!t]
  \centering
 \includegraphics[width=8.5cm,keepaspectratio]{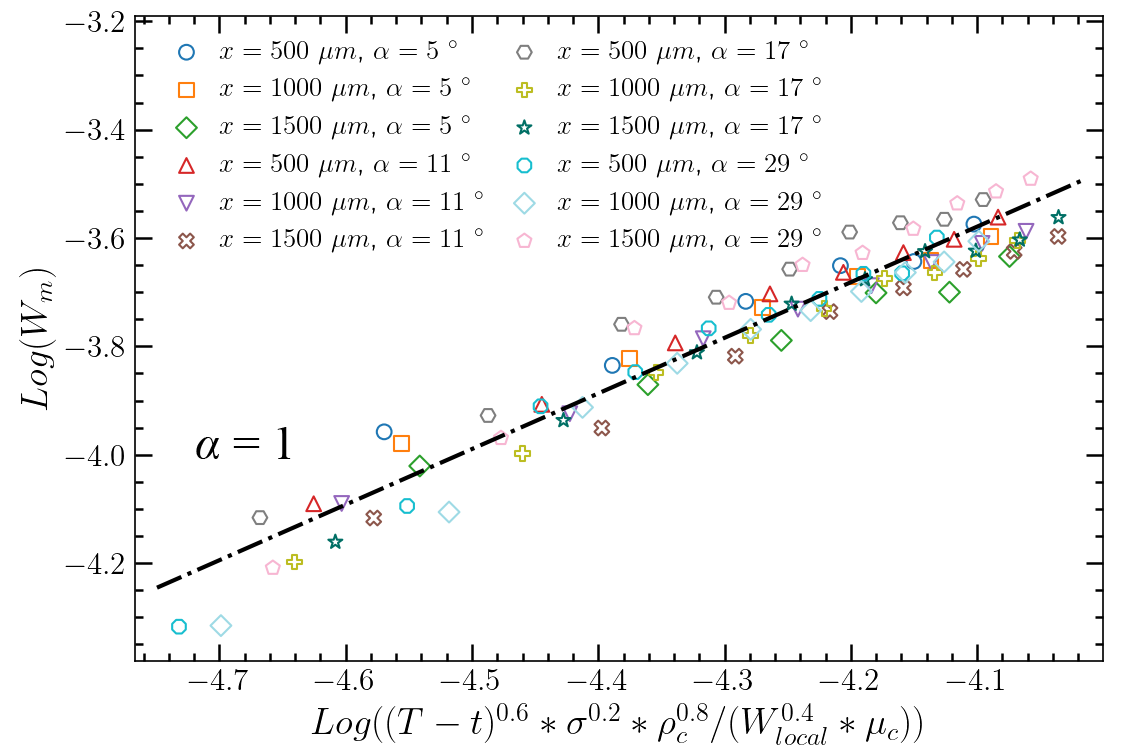}
    \caption{Predictive model for bubble necking width \( W_\text{m} \).}
  \label{fig:8-颈缩宽度原始公式}
\end{figure}
By recasting the dimensionless relationships in terms of physical variables, the predictive model for neck width \( W_\text{m} \) is derived, as shown in Fig. \ref{fig:8-颈缩宽度原始公式}, where the dimensional dependencies on continuous phase properties (\( \rho_\text{c}, \mu_\text{c}, \sigma \)), local geometry (\( W_\text{local} \)), and remaining time (\( T-t \)) are explicitly recovered through inverse dimensional analysis.
\begin{equation}
W_\text{m}\sim \frac{(T-t)^{0.6}\rho_\text{c}^{0.2}\sigma^{0.8}}{W_\text{local}^{0.4}\mu_\text{c}}  
\end{equation}
\label{eq:two-variable system6}
This study establishes a dynamic predictive model for bubble necking width in convergent microchannels through dimensional analysis and experimental scaling laws. By screening key physical mechanisms, the model reduces the original multivariate system to an implicit functional expression dependent solely on continuous phase properties (\( \rho_\text{c} \), \( \mu_\text{c} \), \( \sigma \)), local geometric features (\( W_\text{local} \)), and remaining time (\( T-t \)). The dimensionless analysis reveals a necking width evolution governed by \( W_\text{m} \sim (T-t)^{0.6}/W_\text{local}^{0.4} \), demonstrating a localized control mechanism for necking rupture under microchannel confinement: Reduced local channel width (\( W_\text{local}^{-0.4} \)) exponentially accelerates necking through enhanced velocity gradients.

\section{Interface Curve Similarity}

\begin{figure*}[!t]
  \centering
  \includegraphics[width=15cm,keepaspectratio]{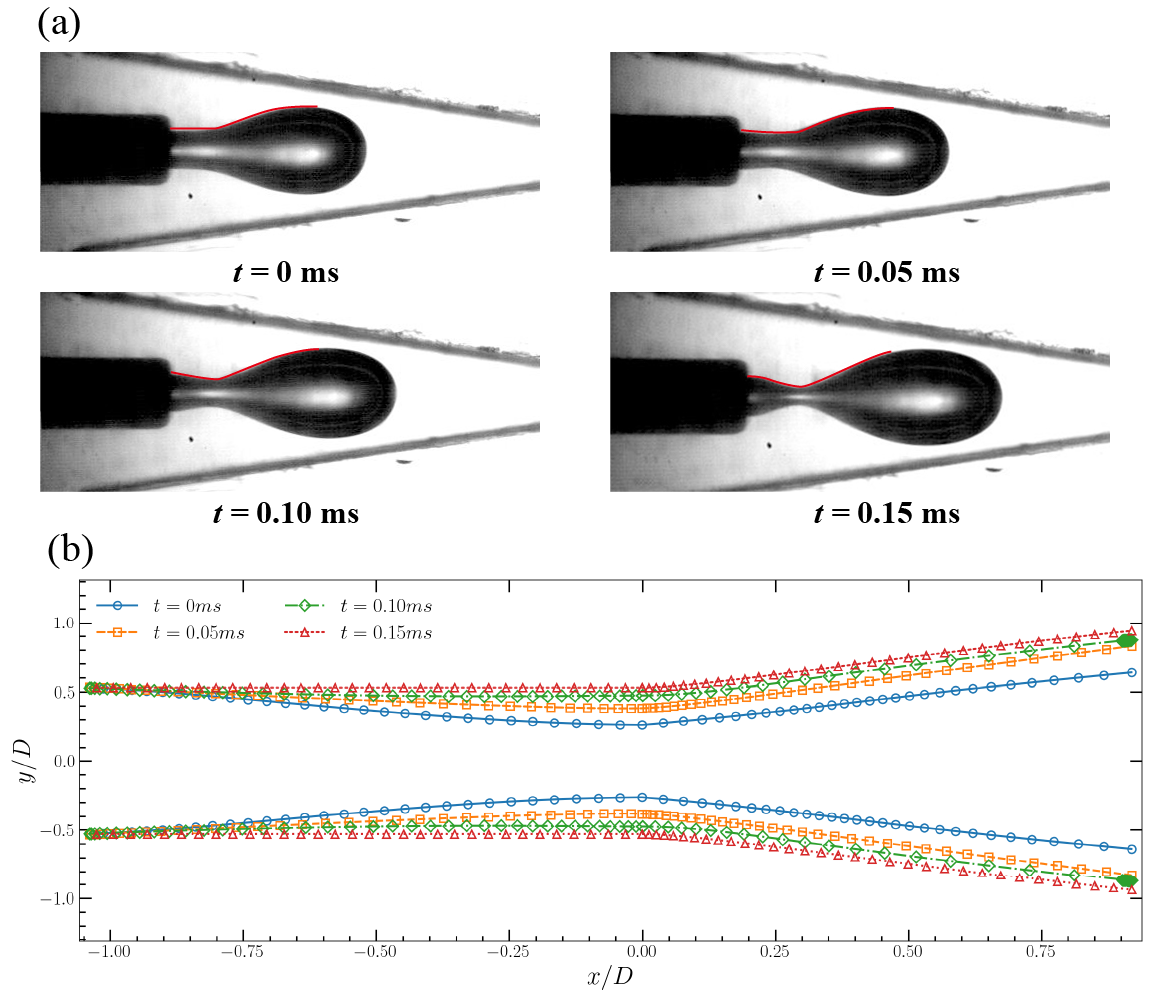}
  \caption{Bubble necking profiles: (a) experimental snapshot; (b) dimensionless position profiles of gas-liquid interface.}
  \label{fig:15-颈缩曲线}
\end{figure*}

While the preceding section investigated the necking width decay process, this section focuses on elucidating the intrinsic correlations governing gas-liquid interface dynamics during necking. Experimental observations reveal that the necking width exhibits specific functional decay characteristics during critical rupture stages, with solution sets demonstrating pronounced self-similar features. The gas-liquid interface curvature solutions in the microchannel convergent region not only exhibit strong coupling with necking width evolution but also share structural similarities with dynamic solutions describing interface collapse processes \cite{zeff2000singularity,lai2018bubble,gordillo2023theory}. This parallels the mathematical formalism governing interface collapse dynamics, motivating systematic characterization of these intrinsic relationships.

Jet interface profiles exhibit universal dimensionless similarity solutions. Zeff et al. \cite{zeff2000singularity} demonstrated through standing wave collapse experiments that as wave height approaches critical value \( h_c \), surface depression generates curvature singularities and triggers jetting, theoretically deriving a self-similar scaling law \( h(r,t)\sim(t_0-t)^{2/3} \) for surface height. Lai et al. \cite{lai2018bubble} further revealed that both cavity collapse during bubble rupture and jet formation obey the inertial-capillary scaling \( \lvert t-t_0\rvert^{2/3} \), consistent with Zeff's exponent \( \alpha=2/3 \). By introducing the Laplace number \( La=\rho\sigma R_0/\mu^2 \), they established the scaling formula \( H(R,T)=T^{2/3}g(RT^{-2/3}) \), which unifies temporal dependencies, fluid properties, and initial radius \( R_0 \), showing collapse of dimensionless data onto a single similarity solution across varying \( La \).

The gas-liquid interface dynamics during necking were captured using a high-speed micro-imaging system. Figure \ref{fig:15-颈缩曲线}(a) illustrates the jet profiles of sample C2 at \( Q_\text{c} = 100 \) ml/min and \( Q_\text{d} = 20 \) ml/min, recorded at \( t = 0 \), 0.05, 0.1, and 0.15 ms. Bubble interface contours were extracted via edge detection algorithms and mapped onto a two-dimensional Cartesian coordinate system: the necking control section’s central axis served as the reference plane (\( x = 0 \)), with the transverse axis aligned along the needle orifice’s axial direction and the longitudinal axis perpendicular to the flow. Spatial normalization was applied by scaling original coordinates (\( x, y \)) with the characteristic length scale \( D \), yielding dimensionless coordinates (Fig. \ref{fig:15-颈缩曲线}(b)). The analysis domain was confined to \( (x/D, y/D) \in (-1, 1) \). Temporal evolution in Fig. \ref{fig:15-颈缩曲线} reveals monotonic reduction of the dimensionless neck width at \( x = 0 \), while the gas-liquid interface geometry progressively elongates with flow development.

\begin{figure*}[!t]
  \centering
  \includegraphics[width=15cm,keepaspectratio]{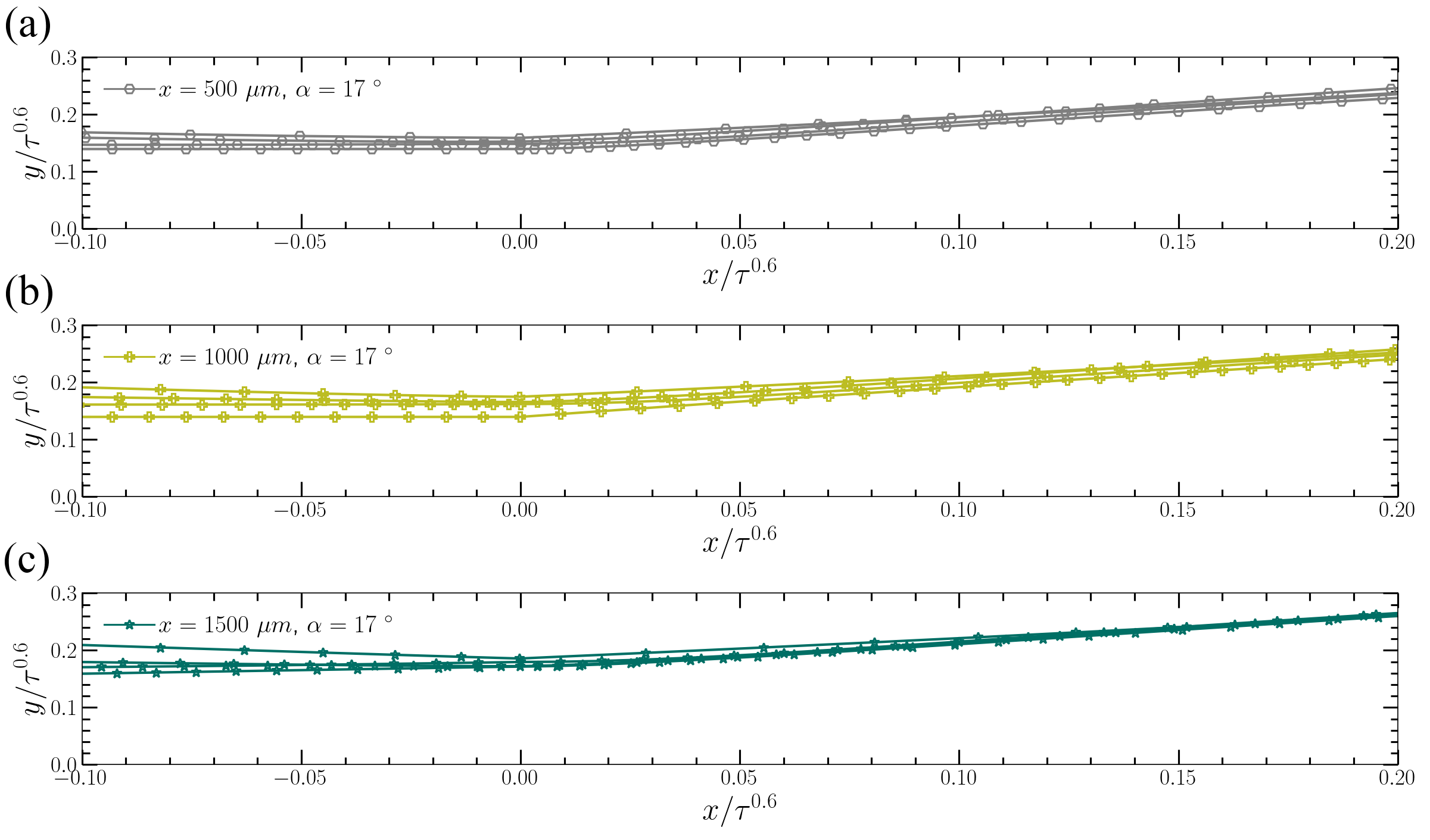}
    \caption{Time-normalized interface curves for different specimens: (a) Specimen C1; (b) Specimen C2; (c) Specimen C3.}
  \label{fig:16-颈缩曲线时间无量纲化}
\end{figure*}

Building upon the jet profile evolution theories of Lai \cite{lai2018bubble} and Zeff \cite{zeff2000singularity}, temporal dynamics were incorporated into the scaling framework by introducing the time scale \((T-t)\). Assuming incompressible and irrotational flow with self-similarity of the gas-liquid interface, the necking curve function was formulated as \( y(x,t) = (T-t)^{\beta_1}f\left(x(T-t)^{-\beta_1}\right) \), where \( y(x,t) \) represents the free surface profile as a bivariate function of spatial coordinate \( x \) and time \( t \), \( f \) denotes the shape function of the interface contour, and \( \beta_1 \) is the temporal scaling exponent. At the necking centerline (\( x = 0 \)), the equation reduces to \( y(t,0) = (T-t)^{\beta_1}f(0) \), where \( f(0) \) acts as a constant prefactor. Given the proportionality \( y(t,0) \sim W_\text{m} \), the necking width follows \( W_\text{m} \sim (T-t)^{\beta_1} \). Experimental observation of the power-law relationship \( W_\text{m} \sim (T-t)^{0.6} \) yields \( \beta_1 = 0.6 \), thereby establishing a dynamical equation with temporal scaling \((T-t)^{0.6}\) and explicit inclusion of the scaling exponent \(\beta_1\).
\begin{equation}
   y(x,t) = (T-t)^{0.6}f(x(T-t)^{-0.6})
    \label{eq: y(x,t)}
\end{equation}
The temporal scale in the equation can be non-dimensionalized as \( \tau = (T-t)\rho_\text{c}\sigma^2/\mu_\text{c}^3 \). Substituting this into Eq. (\ref{eq: y(x,t)}) yields:  
\begin{equation}
   y(x,t) = \tau^{0.6}f(x\tau^{-0.6})
    \label{eq: y(x,tau)}
\end{equation}
Figure \ref{fig:16-颈缩曲线时间无量纲化} presents the dimensionless jet profile functions of samples C1, C2, and C3 under \( Q_\text{c} = 100 \) ml/min, \( Q_\text{d} = 20 \) ml/min, at \( t = 0 \), 0.05, 0.1, and 0.15 ms. By applying Eq. (\ref{eq: y(x,tau)}) to non-dimensionalize the temporal scale of Fig. \ref{fig:15-颈缩曲线}(b), the dimensionless necking curve profiles exhibit significant temporal collapse across all samples, despite variations in their geometric convergence lengths \( x \). This demonstrates that the time-evolving gas-liquid interface necking profiles conform to the self-similar characteristics predicted by Eq. (\ref{eq: y(x,tau)}) under fixed convergence angle conditions.

Figure \ref{fig:16-颈缩曲线时间无量纲化} demonstrates that under fixed convergence angle conditions, dimensionless necking curves of samples with varying convergence lengths exhibit pronounced self-similar characteristics. However, incomplete spatial collapse of the necking profiles persists due to the coupled influence of convergence length \( x \) and convergence angle \( \alpha \). This discrepancy arises from the absence of geometric confinement effects of the microchannel in Eq. (\ref{eq: y(x,t)}), which fails to fully couple spatial constraints into the scaling framework. While temporal normalization in the current model successfully captures interface contraction dynamics, the exclusion of characteristic geometric length scales governing curved interface evolution in convergent microchannels results in systematic deviations between theoretical predictions and experimental data.

\begin{figure*}[!t]
  \centering
  \includegraphics[width=15cm,keepaspectratio]{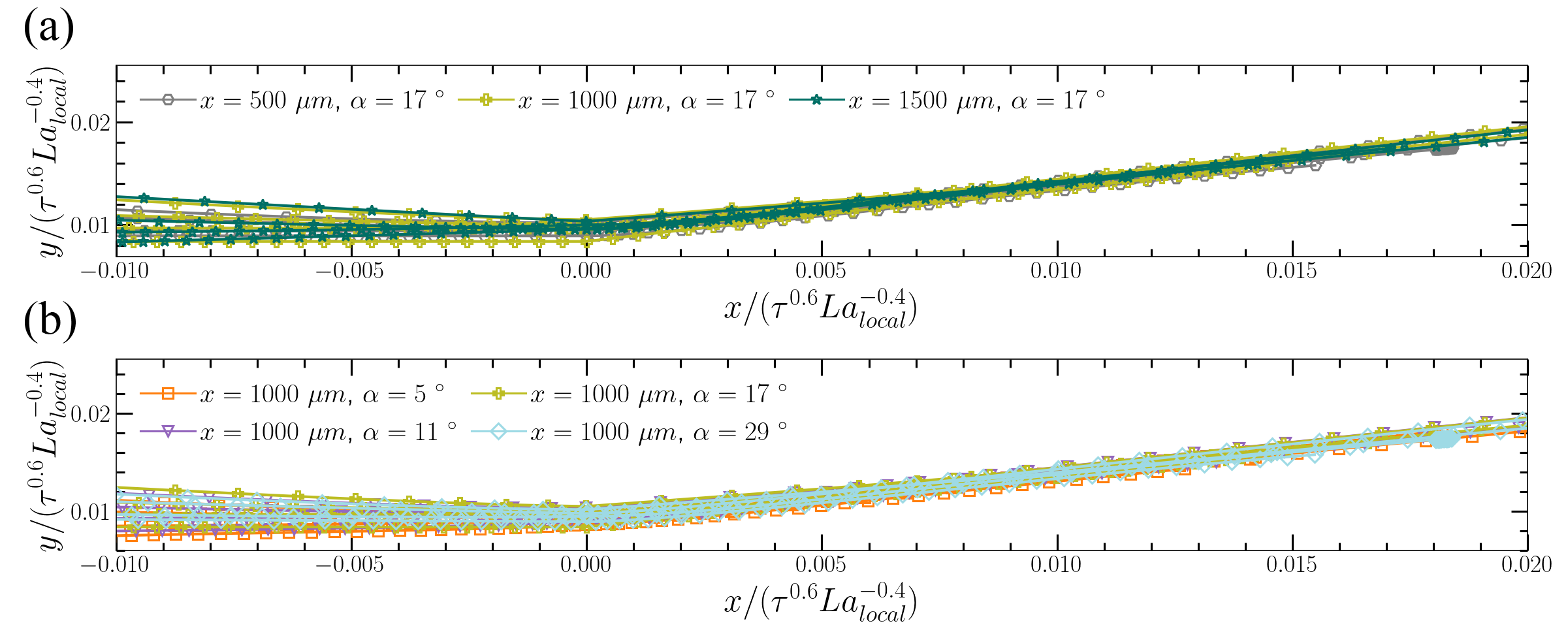}
    \caption{Spatiotemporally normalized interface contour curves for different specimens: (a) Specimens C1, C2, C3; (b) Specimens A2, B2, C2, D2.}

  \label{fig:17-颈缩曲线时间和长度无量纲化}
\end{figure*}

\begin{figure*}[!t]
  \centering
  \includegraphics[width=15cm,keepaspectratio]{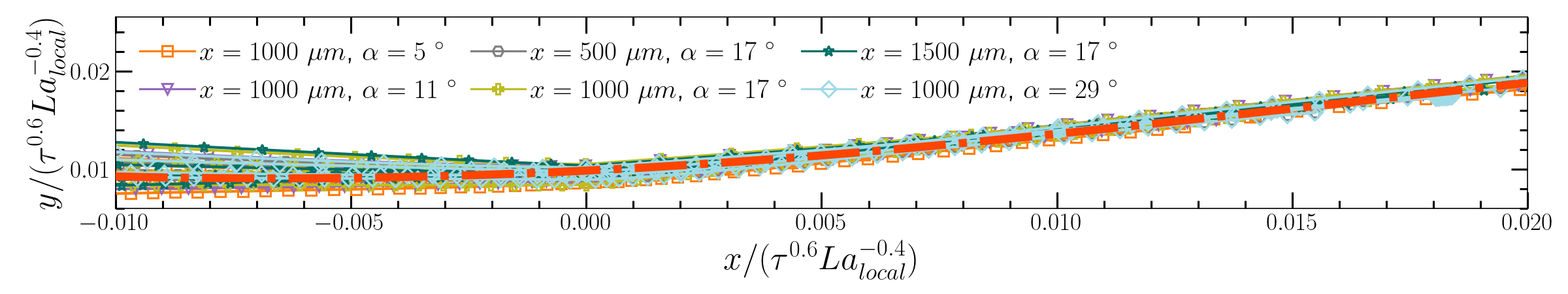}
    \caption{Spatiotemporally normalized interface contour curves for Specimens A2, B2, C1, C2, C3, and D2.}

  \label{fig:18-颈缩曲线数值解}
\end{figure*}

To couple the geometric confinement effects of microchannels, a multi-scale correction is implemented by introducing the local characteristic scale \( W_\text{local} \). Under the assumption of incompressible, irrotational flow and interfacial self-similarity, a modified necking function is constructed: \( y/W_\text{local}^{\beta_2} = \tau^{0.6}f\left(x\tau^{-0.6}/W_\text{local}^{\beta_2}\right) \). This equation employs a dual-parameter scaling law (temporal term \( \tau^{0.6} \) and geometric term \( W_\text{local}^{\beta_2} \)) to simultaneously characterize the spatiotemporal evolution of interface contraction, thereby coupling inertial flow effects with channel geometric constraints. Based on flow similarity, the relationship \( W_\text{m} \sim W_\text{local}^{\beta_2} \) yields \( \beta_2 = -0.4 \), resulting in the fundamental formulation:
\begin{equation}
   \frac{y}{W_\text{local}^{-0.4}} = \tau^{0.6}f(\frac{x\tau^{-0.6}}{W_\text{local}^{-0.4}})
    \label{eq: y(x,t,W_local)}
\end{equation}
The characteristic scale of the convergent channel in the equation can be non-dimensionalized as \( La_\text{local} = W_\text{local}\sigma\rho_\text{c}/\mu_\text{c}^2 \), and substituting this into Eq. (\ref{eq: y(x,t)}) yields a generalized scaling relationship that incorporates both temporal dynamics and geometric confinement effects through the local Laplace number \( La_\text{local} \), enabling unified characterization of interface evolution across varying microchannel geometries:
\begin{equation}
   \frac{y}{La_\text{local}^{-0.4}} = \tau^{0.6}f(\frac{x\tau^{-0.6}}{La_\text{local}^{-0.4}})
    \label{eq: y(x,t,La_local)}
\end{equation}
Figure \ref{fig:17-颈缩曲线时间和长度无量纲化}(a) presents the spatiotemporally dimensionless gas-liquid interface necking functions of samples C1, C2, and C3 under \( Q_\text{c} = 100 \) ml/min and \( Q_\text{d} = 20 \) ml/min at \( t = 0 \), 0.05, 0.1, and 0.15 ms, representing cases with identical convergence angle \( \alpha \) but varying convergence lengths \( x \). Figure \ref{fig:16-颈缩曲线时间无量纲化}(b) displays the jet profile functions of samples A2, B2, C2, and D2 under the same flow rates and temporal conditions, corresponding to cases with fixed convergence length \( x \) but differing convergence angles \( \alpha \). The results in Fig. \ref{fig:17-颈缩曲线时间和长度无量纲化} demonstrate high universality of necking curves in convergent microchannels, where the spatiotemporal evolution of interface contraction aligns with the self-similar characteristics predicted by Eq. (\ref{eq: y(x,t,La_local)}). This validates that the theoretical model, through coupling the local characteristic scale \( W_\text{local} \), effectively isolates the influence of microchannel geometric parameters on interface dynamics.

Figure \ref{fig:18-颈缩曲线数值解} displays the spatiotemporally dimensionless interface profile functions of gas-liquid necking curves for samples A2, B2, C1, C2, C3, and D2 under \( Q_\text{c} = 100 \) ml/min, \( Q_\text{d} = 20 \) ml/min, at \( t = 0 \), 0.05, 0.1, and 0.15 ms. Curve fitting was performed using machine learning-based polynomial regression with a stepwise order selection strategy: the data were randomly partitioned into training and validation sets (8:2 ratio), and optimal model selection within 1st- to 5th-order polynomials was automated based on the validation set coefficient of determination (\( R^2 = 0.93 \)). The fitted curve (red dashed line in Fig. \ref{fig:18-颈缩曲线数值解}) enables explicit determination of the mathematical expression for Eq. (\ref{eq: y(x,t,La_local)}) (rounded to two decimal places):
\begin{equation}
   y' = 0.01 + 0.23x'^1 + 16.49x'^2 - 158.79x'^3 - 6475.72x'^4
    \label{eq: y'(x')}
\end{equation}
Here, \( y' = y/(La_\text{local}^{-0.4}\tau^{0.6}) \) and \( x' = x/(La_\text{local}^{-0.4}\tau^{0.6}) \) represent the dimensionless forms of the \( x \)- and \( y \)-coordinates for the bubble necking curve. Equation (\ref{eq: y'(x')}) demonstrates that the necking interface profiles exhibit spatiotemporal self-similarity and conform to the flow similarity in convergent microchannels \cite{wang2015speed}. Comparative analysis between the present gas-liquid interface model and existing jet-cavity models reveals a critical distinction: while traditional frameworks rely on feature radius \( R \)-dominated governance of interface evolution, the proposed necking curve model explicitly depends on the local Laplace number \( La_\text{local} \), marking a fundamental divergence from established theories.

\section{Summary and conclusions}

Experimental observations reveal that monodisperse bubble generation exhibits four distinct stages: growth, necking, detachment, and stabilization. By constructing a bubble necking rupture model and integrating dimensional analysis with experimental data, this study systematically investigates the influence of nozzle convergence length \( x \) and convergence angle \( \alpha \) on bubble necking dynamics, establishing a scaling relationship for neck width \( W_\text{m} \) with remaining time \( (T-t) \) and characteristic scale \( W_\text{local} \): \( W_\text{m} \sim \left[(T-t)^{0.6}\rho_\text{c}^{0.2}\sigma^{0.8}\right]/(W_\text{local}^{0.4}\mu_\text{c}) \). Experiments demonstrate that under identical flow rates, the power-law scaling of \( W_\text{m} \) remains consistent across varying geometric configurations, exhibiting robust universality. For shear-dominated bubble necking processes, the temporal evolution of necking interface profiles adheres to the \( (T-t)^{0.6} \) power-law scaling, while spatial distributions follow the \( W_\text{local} \)-dependent scaling. A machine learning polynomial regression approach, employing an 8:2 training-validation split and stepwise 1st- to 5th-order model selection, yields the optimal mathematical expression: \( y' = 0.01 + 0.23x'^1 + 16.49x'^2 - 158.79x'^3 - 6475.72x'^4 \). This model achieves unified self-similar characterization of necking curves under diverse spatiotemporal conditions, elucidating the universal spatiotemporal evolution laws governing bubble necking interface dynamics.


\section*{Acknowledgements}
This work was supported by the National Natural Science Foundation of China-Yunnan Joint Fund Key Project, U2002214. The authors would like to express their sincere gratitude to Professors Jin-Song Zhang, Guo-Hui Hu, Zhe-Wei Zhou, and Qian Xu for valuable assistance.

\bibliography{apssamp}

\end{document}